\documentclass[
 longbibliography,
 twocolumn,
 superscriptaddress,
 amsmath,
 amssymb,
 amsthm,
 amsfonts,
 aip,
 floatfix,
 reprint
]{revtex4-2}

\usepackage[colorlinks=true,urlcolor=blue,citecolor=blue,linkcolor=blue]{hyperref} 
\usepackage{graphicx}
\usepackage{dcolumn}
\usepackage{kotex}
\usepackage{color}
\usepackage{amsfonts}
\usepackage{hyperref}
\usepackage{bm}
\usepackage{bbm}

\definecolor{c}{rgb}{0,0.6,0.6}
\definecolor{m}{rgb}{0.6,0,0.6}
\definecolor{y}{rgb}{0.6,0.6,0}

\begin{document}

\title{Unexpected advantages of exploitation for target searches in complex networks}

\author{Youngkyoung Bae}
 \affiliation{Department of Physics, Korea Advanced Institute of Science and Technology, Daejeon 34141, Korea}
 
\author{Gangmin Son}
 \affiliation{Department of Physics, Korea Advanced Institute of Science and Technology, Daejeon 34141, Korea}

\author{Hawoong Jeong}
\email{hjeong@kaist.edu}
\affiliation{Department of Physics, Korea Advanced Institute of Science and Technology, Daejeon 34141, Korea}
\affiliation{Center of Complex Systems, Korea Advanced Institute of Science and Technology, Daejeon 34141, Korea}

\date{\today}

\begin{abstract}

Exploitation universally emerges in various decision-making contexts, e.g., animals foraging, web surfing, the evolution of scientists' research topics, and our daily lives.
Despite its ubiquity, exploitation, which refers to the behavior of revisiting previous experiences, has often been considered to delay the search process of finding a target.
In this paper, we investigate how exploitation affects search performance by applying a non-Markovian random walk model, where a walker randomly revisits a previously visited node using long-term memory.
We analytically study two broad forms of network structures, namely (i) clique-like networks and (ii) lollipop-like networks, and find that exploitation can significantly improve search performance in lollipop-like networks whereas it hinders target search in clique-like networks.
Moreover, we numerically verify that exploitation can reduce the time needed to fully explore the underlying networks by using $550$ diverse real-world networks.
Based on the analytic result, we define the \textit{lollipop-likeness} of a network and observe a positive relationship between the advantage of exploitation and lollipop-likeness.

\end{abstract}

\maketitle

\begin{quotation}
Why is exploitation ubiquitous in diverse decision-making situations?
Does exploitation help achieve quick target search?
The intuitive answer is no, meaning that exploiting previously visited places squanders time in wrong places, thereby degrading search performance.
Counterintuitively, using a non-Markovian random walk model, we show here that exploitation can in fact help target search in complex networks. Our random walk process is performed by randomly deciding to either explore any neighbor of the current node or exploit a previously visited node.
We analytically reveal that exploitation benefits target search in lollipop-like but not clique-like networks.
Demonstrating the improved results of our model with $550$ real-world networks, this work provides a clue in answering why many organisms frequently exploit their known areas and suggests a new direction in the development of efficient searching algorithms that can be applied in diverse fields such as computer science, sociology, etc.

\end{quotation}

\section{Introduction}
\label{sec:intro}

\begin{figure}[!b]
    \includegraphics[width=\columnwidth]{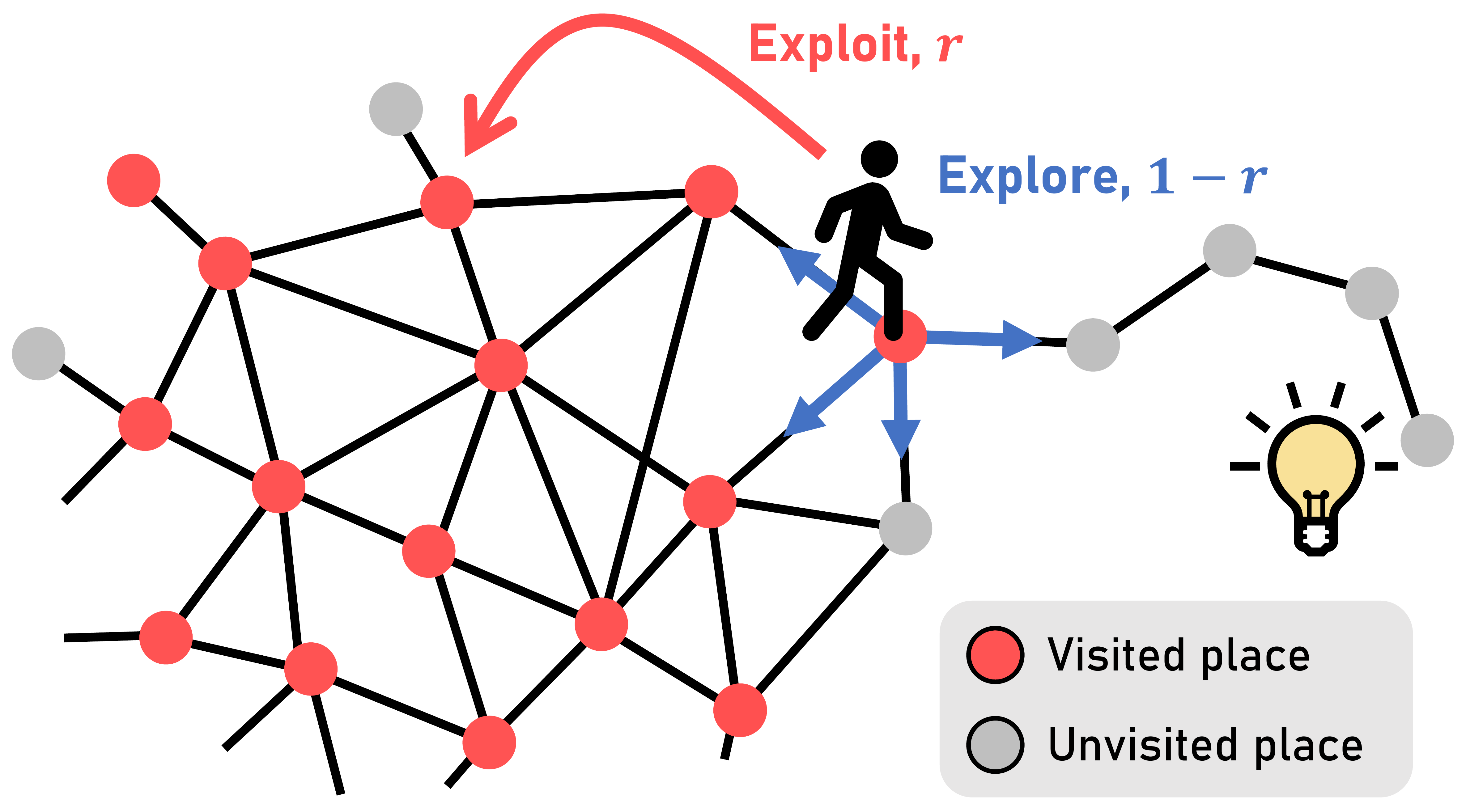}
    \vskip -0.1in
    \caption{ Schematic of a stochastic exploiting random walk (SERW) in a network. A walker either explores any adjacent node with probability $1-r$ or exploits one of their previously visited nodes (red nodes) with probability $r$. When a walker seeks to make a discovery, will exploitation help or hinder their goal?
    }\label{fig1}
\end{figure}

A strategic tension between exploration and exploitation emerges as a universal phenomenon in a diverse range of decision-making contexts~\cite{march1991exploration, gupta2006interplay, cohen2007should}.
In general, the concept of exploration refers to the behavior of seeking new possibilities, whereas exploitation refers to the behavior of revisiting previous experiences (see Fig.~\ref{fig1}).
Animal mobility patterns are typical examples of the tension between the two, revealing a switching behavior between wandering in a search of new areas and revisiting familiar places~\cite{krebs1978test, boyer2014random, merkle2014memory, hooten2017animal}.
Human movements in physical space and cyberspace have also been commonly modeled by including the two strategies~\cite{song2010modelling, zhao2014scaling}.
In addition, much attention has recently been paid to the research-interest evolution of scientists, or in other words their mobility in abstract knowledge space.
Before a scientist chooses their next topic, he or she is torn between adhering to familiar research in their domain (productive tradition) and challenging the acquisition of new knowledge (risky innovation)~\cite{bourdieu1975specificity, kuhn1977essential, foster2015tradition, jia2017quantifying, iacopini2018network,zeng2019increasing, liu2021understanding}.
To understand the implications of these prevalent patterns, it is essential to unveil their strategic benefits.

Despite the ubiquity of exploration/exploitation tension, in some problems or organizations, it is more appropriate to solely utilize exploration (or exploitation) without balance depending on the objectives.
One such problem is the random target search problem, where a random walker seeks to visit a particular target place or all places.
The widespread belief in this problem is that exploiting previously visited places wastes time by needless actions and thus impedes the quick search.
This negative relation seems quite intuitive, and numerous search strategies have been constructed in favor of the relation~\cite{madras1993self, yang2005exploring, kim2016network, arruda2017knowledge, lima2018dynamics}. 
However, recent studies have theoretically revealed that target search with returns can reduce the search time~\cite{evans2011diffusion, riascos2020random, pal2020search}, which confuses the known relationship between exploration and exploitation. 
Therefore, our main questions naturally arise: how does exploitation with long-term memory affect search performance? How does this implication vary for complex search spaces (i.e., networks)?

%%%%%%%%%%%%% figure 1 %%%%%%%%%%%%%%
%%%%%%%%%%%%% figure 1 %%%%%%%%%%%%%%

Specifically, we investigate the effects of exploitation on target search with a proposed model called the stochastic exploiting random walk (SERW), where a walker decides either to explore any of its neighboring nodes or to exploit a previously visited node.
Random walks have been extensively adopted as a general framework and basically modeled as Markovian~\cite{pearson1905problem, codling2008random, masuda2017random}; nevertheless, there have been empirically reported non-Markovian effects in numerous systems~\cite{song2010modelling, guerin2012nonmarkovian, fagan2013spatial}.
A few non-Markovian models have been studied~\cite{schutz2004elephants, boyer2014random, kim2016self, falcon2017localization}, but the scarcity of their analytical results has hampered our understanding of the related memory effects.
Here, we analytically solve the SERW with long-term memory in terms of search performance in two extreme structures: a clique-like structure and a lollipop-like structure.
We show that the exploitation process can be beneficial in the lollipop-like networks, while it does not improve the search performance in the clique-like networks.
In addition, we propose a measure, \textit{lollipop-likeness}, that quantifies how close a particular network is to having the lollipop structure, or alternatively, how far it is from the clique structure.

The paper is organized as follows. In Sec.~\ref{sec:serw}, we analytically investigate the search performance of the SERW in a clique-like network and a lollipop-like network.
To quantify search performance, we employ the measures first-passage time (FPT) and cover time (CT), defined as the time needed to find a single target node or cover the entire network, respectively~\cite{redner2001aguide}.
Between them, the FPT-related quantities have been extensively adopted in studies of the search performance for a single target node~\cite{noh2004random, condamin2007first, guerin2016mean}.
On the other hand, while the CT-related quantities have been studied for characterizing how fast a walker covers the entire underlying network, analytical and heuristic results from this measure are uncommon~\cite{characteristic2014bonaventura, chupeau2015cover, maier2017cover}.
Based on these quantities, we show under what conditions exploitation can be beneficial in the search process (see Table~\ref{tab1} for the list of abbreviations for FPT and CT-related quantities).
In Sec.~\ref{sec:real_networks}, we observe that exploitation can significantly reduce the time to cover $550$ different real-world networks by simulations. 
Finally, we conclude our work in Sec.~\ref{sec:conclusions}.

\section{Stochastic exploiting random walk and analytical results}
\label{sec:serw}

We consider a discrete-time random walk in an undirected, unweighted, and connected network of $N$ nodes labeled by $i=0,1, \dots, N-1$. 
Elements of an adjacency matrix $\mathsf{A}$ satisfy $\mathsf{A}_{ij}=\mathsf{A}_{ji}$, where $\mathsf{A}_{ij}=1$ if two nodes $i$ and $j$ are linked or $\mathsf{A}_{ij}=0$ otherwise. 
At each time step $t$, a walker decides the process of either exploration or exploitation with probability $1-r$ and $r$, respectively.
The exploration process corresponds to the normal random walk (NRW), i.e., the walker explores one of the adjacent nodes and succeeds (fails) to achieve a discovery if the chosen node has been unvisited (visited) before. 
A transition probability from $i$ to $j$ in the exploration process is defined by $\mathsf{W}_{ij} = \mathsf{A}_{ij}/k_i$ where $k_i=\sum_{j=0}^{N-1}\mathsf{A}_{ij}$ is the degree of a node $i$ and $\mathsf{W}$ is called the transition matrix~\cite{noh2004random}. 
By contrast, exploitation is implemented as a process where the walker randomly jumps to one of the previously visited nodes. 
We set that all visited nodes have the same probability to be chosen in this process for simplicity. 
Note that the SERW model becomes equivalent to the NRW when $r=0$.

The mean first-passage time (MFPT) $T_{ij}\left( r | \mathcal{S} \right)$, defined as the average time needed to visit $j$ from $i$ by the SERW with $r$, follows the backward equation:
\begin{equation}\label{eq:SERW}
\begin{aligned}
    T_{ij}\left(r | \mathcal{S}\right) = (1-r) \sum_{l=0}^{N-1}\mathsf{W}_{il} T_{lj} \left(r | \mathcal{S}'\right) \\
    + \frac{r}{|\mathcal{S}|}\sum_{l \in \mathcal{S}} T_{lj}\left( r|\mathcal{S} \right) + 1,
\end{aligned}
\end{equation}
where $\mathcal{S}$ is a set of distinct nodes used for the exploitation process, $|\mathcal{S}|$ is the cardinality of $\mathcal{S}$, and $\mathcal{S}' \equiv \mathcal{S} \cup \{l\}$ is the updated set by the exploration process; see the derivation in Appendix~\ref{sec:AppendixA}.
At $t=0$, the walker starts with $\mathcal{S}$ including only an initial node. 
When the walker explores its neighbor node, $\mathcal{S}$ is updated to $\mathcal{S}'$ [the first term in Eq.~\eqref{eq:SERW}]; otherwise, the walker cannot visit any new node in the exploitation process, so $\mathcal{S}$ has no update [the second term in Eq.~\eqref{eq:SERW}]. 
Our model has similarities with the stochastic resetting random walks~\cite{evans2011diffusion, riascos2020random, pal2020search, gonzalez2021diffusive, wang2021random}, but the update rule of $\mathcal{S}$ over a trajectory creates discrepancies including path dependence.
We also consider the global mean first-passage time (GMFPT) for a target node $j$ defined by the average of the MFPT over all possible starting nodes except for the target node~\cite{tejedor2009global}:
\begin{equation}\label{eq:GMFPT_def}
\begin{aligned}
    T_{j}\left(r | \mathcal{S} \right) =   \frac{1}{N-1} \sum_{i \neq j} T_{ij}(r|\mathcal{S}).
\end{aligned}
\end{equation}
By definition, the GMFPT allows us to estimate how long a walker will take to reach the target node from a randomly selected starting node.

The mean cover time (MCT) $C_i(r)$ is defined here as the average time needed to visit all nodes at least once from a starting node $i$ by the SERW with $r$.
Some analytical results of the MCT of NRW or other random walks have been studied using lattices, Erd\H{o}s--R\'enyi (ER) networks, Barab\'asi--Albert (BA) networks, etc~\cite{yokoi1990some, cooper2007cover_ER, cooper2007cover_BA, chupeau2015cover, maziya2020dynamically}.
For any connected network, it is well-known that the MCT of NRW $C_i^{\rm NRW}$ is bounded as~\cite{feige1995tight_upper, feige1995tight_lower}
\begin{equation}\label{eq:cover_bound}
\begin{aligned}
    \left( 1 + o(1) \right)N \ln N \leq C_i^{\rm NRW} \leq \left( 1 + o(1) \right)\frac{4}{27}N^3,
\end{aligned}
\end{equation}
for any node $i$, where $o(1)$ denotes a term that converges to $0$ as $N$ increases.
The superscript `${\rm NRW}$' indicates the value for the NRW.
The lower bound and the upper bound of $C_i^{\rm NRW}$ can be obtained in a clique network and a lollipop network, respectively, the latter of which is composed of a chain attached to a clique (see Appendix~\ref{sec:AppendixB} and \ref{sec:AppendixC}).
To focus on the areas around these bounds, in the following subsections, we analyze structures close to these two extreme cases, which we call clique-like networks and lollipop-like networks.
The global mean cover time (GMCT) $C(r)$ is defined by $\sum_{i=0}^{N-1}{C_i(r)}/N$. 
In this paper, $C(r)$ is mainly applied to verify how fast a walker uncovers entire networks in simulations in Sec.~\ref{sec:real_networks}.

\begin{table}[!b]
    \centering
    \caption{List of abbreviations}
    \label{tab1}
    \begin{tabular}{c|c}
    \noalign{\smallskip}\noalign{\smallskip}\hline\hline
    Abbreviation & Definition \\
    \hline
    SERW & stochastic exploiting random walk \\
    NRW & normal random walk \\
    FPT & first-passage time \\
    CT & cover time \\
    (G)MFPT & (global) mean first-passage time \\
    (G)MCT & (global) mean cover time \\
    ER network  & Erd\H{o}s--R\'enyi network  \\
    BA network  & Barab\'asi--Albert network  \\
    \hline
    \hline
    \end{tabular}
\end{table}

\subsection{Clique-like networks}

In clique-like networks, a walker can reach any node with small steps, and the steady-state distribution is rapidly achieved by a short relaxation time $t_{\rm rlx}$. 
When a walker wanders a network with $t_{\rm rlx} \ll N$, the information correlated to the starting node rapidly vanishes~\cite{lau2010asymptotic}. 
This fact implies that the revisited nodes by the exploitation process hardly affect the exploration process except for wasting time by useless steps when $r \ll 1$.
Thus, $T_j(r)$ is approximately given by
\begin{equation}\label{eq:GMFPT_1}
\begin{aligned}
    T_{j}\left(r \right) \simeq \frac{T_{j}^{\rm NRW}}{1-r}.  \\
\end{aligned}
\end{equation}
Note that Eq.~\eqref{eq:GMFPT_1} becomes less accurate as $r$ increases because the information of the revisited nodes does not sufficiently vanish at large $r$.
For a clique network, which is the extreme case of our clique-like networks, it can be easily derived that the equality is exactly satisfied by Eq.~\eqref{eq:SERW} (Appendix~\ref{sec:AppendixB}). 
$T_j^{\rm NRW}$ for an arbitrary network can be calculated by using the graph Laplacian $\mathsf{L}$ defined by $\mathsf{L}_{ij} = \delta_{ij}k_i - \mathsf{A}_{ij}$ with Kronecker's delta $\delta_{ij}$~\cite{lin2012mean}:
\begin{equation}\label{eq:GMFPT_NRW}
\begin{aligned}
    T_{j}^{\rm NRW} = \frac{N}{N-1}\sum_{l=1}^{N-1} \frac{1}{\mu_l} \left( 2E\psi_{lj}^2 - \psi_{lj}\sum_{z=1}^{N} k_z \psi_{lz} \right).  \\
\end{aligned}
\end{equation}
Here, the total number of edges is $E=\sum_{z=0}^{N-1}k_z/2$, $\mu_l$ is the eigenvalue of $\mathsf{L}$ with $0=\mu_0 \leq \mu_1 \leq \dots \leq \mu_{N-1}$, and $\bm{\psi}_l = (\psi_{l0}, \dots, \psi_{l, N-1})^{\top}$ is the corresponding eigenvector of unit length.

%%%%%%%%%%%%% figure 2 %%%%%%%%%%%%%%
\begin{figure}[!t]
    \includegraphics[width=\columnwidth]{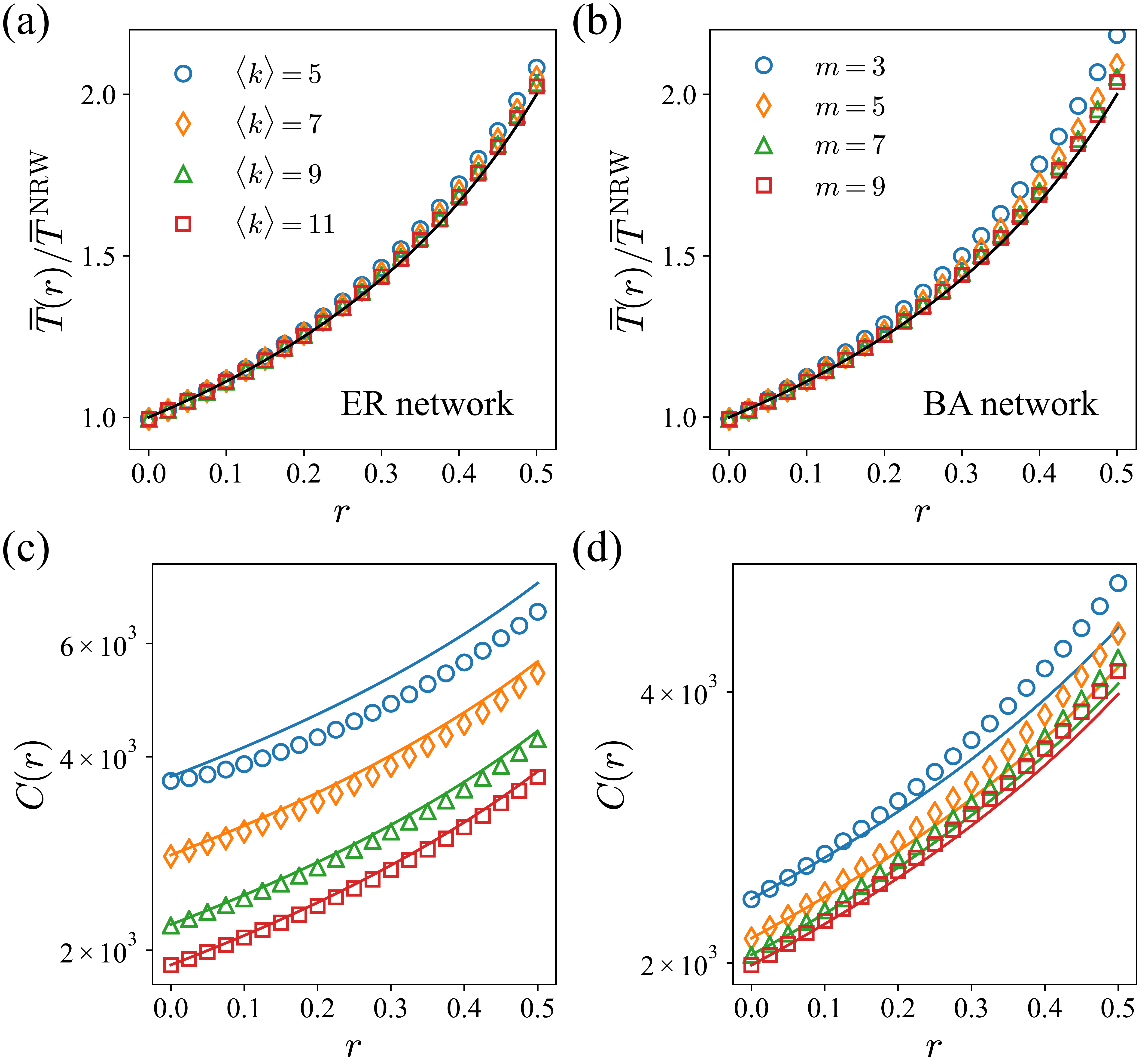}
    \vskip -0.1in
    \caption{ (a,b) Ratio between the mean GMFPT of SERW and NRW denoted by $\overline{T}(r)/\overline{T}^{\rm NRW}$, as a function of the exploit probability $r$ in Erd\H{o}s--R\'enyi (ER) (a) and Barab\'asi--Albert (BA) (b) networks.
    (c,d) $C(r)$ as a function of $r$ in ER (c) and BA (d) networks.
    All results are examined for $100$ realizations with $N=200$ and different parameters, namely the mean degree $\langle k \rangle$ and the number of new links per node $m$ for ER and BA networks, respectively.
    Simulations are performed $25$ runs per node (i.e., $5000$ runs) with each network.
    Solid lines (symbols) indicate the analytical (simulation) results.
    }\label{fig2}
\end{figure}
%%%%%%%%%%%%% figure 2 %%%%%%%%%%%%%%

%%%%%%%%%%%%% figure 3 %%%%%%%%%%%%%%
\begin{figure*}[!t]
    \includegraphics[width=\linewidth]{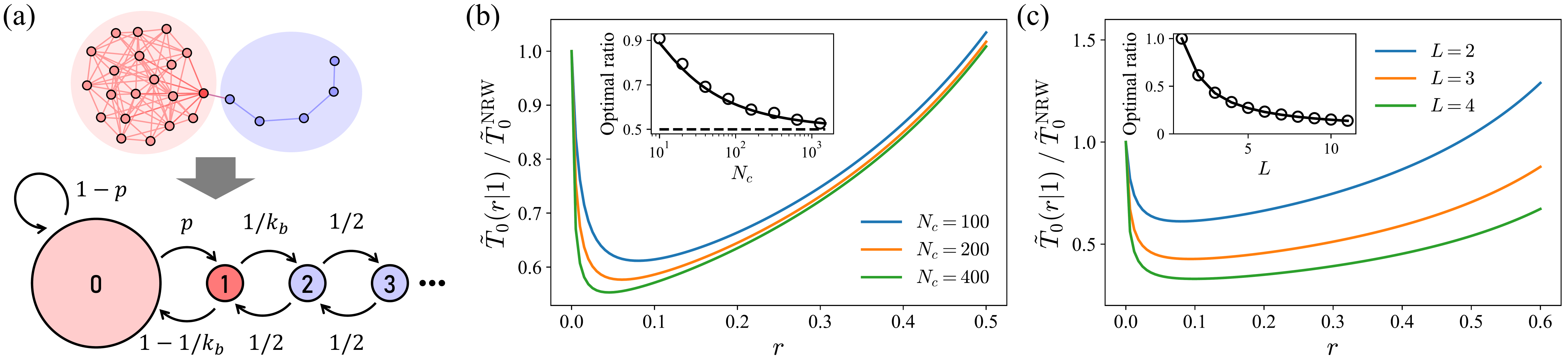}
    \vskip -0.1in
    \caption{ (a) Illustrations of (top) a lollipop-like network and (bottom) its coalesced version. The supernode $0$ indicates the coalesced core with $N_c$ nodes, node $1$ indicates a bridge node with degree $k_b$, and nodes $2, \dots, L+1$ indicate the chain with length $L$.
    (b,c) Ratio between the MFPTs of the SERW and the NRW, denoted by $\tilde{T}_0(r|1)/\tilde{T}_0^{\rm NRW}$, as a function of the exploit probability $r$ with different $N_c$ and $L$ at $L=2$ and $N_c=100$, respectively.
    Here, we set the core as a clique network, so $p=1/N_c$ and $k_b=N_c+1$.
    Insets show the optimal ratio $\tilde{T}_0(r^{*}|2)/\tilde{T}_0^{\rm NRW}$ as a function of (b) $N_c$ and (c) $L$, obtained at $r^*$ which minimizes $\tilde{T}_0(r|1)$. In the insets, the dashed line indicates the asymptotic value of the optimal ratio, and the solid lines (symbols) indicate the analytical (simulation) results.
    }\label{fig3}
\end{figure*}
%%%%%%%%%%%%% figure 3 %%%%%%%%%%%%%%

We next study $C(r)$ as a function of $r$. 
Here, $C(r)$ is equal to $C_i(r)$ since the starting node $i$ does not influence $C_i(r)$.
$C^{\rm NRW}$ with $t_{rlx} \ll N$ can be computed by using the FPT statistics, derived in Ref.~\onlinecite{maier2017cover}, and we follow the same process with the inclusion of $r$ to obtain $C(r)$.
The FPT distribution $F_{ij}(t, r|\mathcal{S})$ for a target node $j$ from a starting node $i$ in a network with $t_{\rm rlx} \ll t$ exponentially decays as follows~\cite{kittas2008trapping, lau2010asymptotic}:
\begin{equation}\label{eq:FPT_exp(t)}
\begin{aligned}
    F_{ij}(t, r|\mathcal{S}) = A_j(r) \exp{\left[-t/T_j(r) \right]},
\end{aligned}
\end{equation}
with $F_{ij}(0, r|\mathcal{S})=\delta_{ij}$. 
Normalizing $F_{ij}(t, r|\mathcal{S})$ from $t=1$ to $t=\infty$ with discrete-time $t$, we can determine that $A_j(r)=\exp{\left[1/T_j(r) \right]}-1$.
Then, the cumulative distribution of FPT $\mathcal{P}(t' \leq t )$ is calculated by
\begin{equation}\label{eq:FPT_cum(t)}
\begin{aligned}
    \mathcal{P}\left(t' \leq t \right) &= \sum_{t'=1}^{t} A_j(r) \exp{\left[-t'/T_j(r) \right]}
    \\&= 1-\exp{\left[- t/T_j(r) \right]}.
\end{aligned}
\end{equation}
The cumulative distribution of the cover time, denoted by $\rho(t)$, is equivalent to the probability that all FPTs are less than or equal to $t$, yielding
\begin{equation}\label{eq:rho(t)}
\begin{aligned}
    \rho(t) &= \prod_{i\neq j}\Big( 1-\exp{\left[-t/T_j(r) \right]} \Big).
\end{aligned}
\end{equation}
Therefore, $C(r)$ can be obtained by
\begin{equation}\label{eq:C(r)}
\begin{aligned}
    C(r) = \sum_{t=1}^{\infty} [1-\rho(t)]
    \simeq \int_{0}^{\infty} dt \left[ 1-\rho(t) \right].
\end{aligned}
\end{equation}
We can approximate $t$ as a continuous variable in the last term due to a negligibly small error by this change compared to $C(r)$~\cite{maier2017cover}.
Note that $T_j(r)$ is larger than $T^{\rm NRW}$, making $\rho(t)$ slowly converge to $1$ as $r$ increases.
As a result, we can obtain that $C(r)$ also increases with increasing $r$ in the clique-like networks.

We examine whether Eqs.~\eqref{eq:GMFPT_1} and \eqref{eq:C(r)} are valid for ER and BA networks with small relaxation times belonging to clique-like networks.
In this paper, $\langle \cdot \rangle$ is the ensemble average, $\langle k \rangle$ is the average degree, and $m$ is the number of new edges at node creation.
To verify the relation $T_j(r)/T_j^{\rm NRW} \simeq 1/(1-r)$, we numerically obtain the mean GMFPT $\overline{T}(r)$ defined by the average of the GMFPTs over all possible target nodes.
As depicted in Fig.~\ref{fig2}(a) and (b), Eq.~\eqref{eq:GMFPT_1} for both networks agrees well even until $r=0.5$ despite their different degree distributions.
Our predictions become more precise as $\langle k \rangle$ and $m$ increase, because the relaxation time decreases from $\langle t_{\rm rlx}\rangle =5.17$ ($3.47$) to $\langle t_{\rm rlx}\rangle =2.27$ ($1.78$) as $\langle k \rangle$ ($m$) increases.
Similarly, Eq.~\eqref{eq:C(r)} holds well until $r=0.5$ and becomes more consistent with the simulation results as $\langle k \rangle$ and $m$ increase, as shown in Fig.~\ref{fig2}(c) and (d).
% We examine Eqs.~\eqref{eq:GMFPT_1} and \eqref{eq:C(r)} for ER and BA networks with different parameters, as shown in Fig.~\ref{fig2}.
% In this paper, $\langle \cdot \rangle$ is the ensemble average, $\langle k \rangle$ is the average degree, and $m$ is the number of new edges at node creation.
% To verify the relation $T_j(r)/T_j^{\rm NRW} \simeq 1/(1-r)$, we numerically obtain the mean GMFPT $\overline{T}(r)$ defined by the average of the GMFPTs over all possible target nodes.
% As shown in Fig.~\ref{fig2}(a) and (b), Eq.~\eqref{eq:GMFPT_1} for both networks agrees well even until $r=0.5$.
% Our predictions become more precise as $\langle k \rangle$ and $m$ increase, because the relaxation time decreases from $\langle t_{\rm rlx}\rangle =5.17$ ($3.47$) to $\langle t_{\rm rlx}\rangle =2.27$ ($1.78$) as $\langle k \rangle$ ($m$) increases.
% Similarly, Eq.~\eqref{eq:C(r)} holds well until $r=0.5$ and becomes more consistent with the simulation results as $\langle k \rangle$ and $m$ increase, as shown in Fig.~\ref{fig2}(c) and (d).

\subsection{Lollipop-like networks}

We consider a lollipop-like network consisting of two parts, a core and a chain as illustrated in Fig.~\ref{fig3}(a), analogous to the candy and stick of a lollipop.
Since the approximation we used in the previous section is not valid for a lollipop-like network, we propose a coarse-graining approach.
Assuming that the core is already covered while the chain is not because the core has much shorter GMFPTs, we represent the core with $N_c$ nodes as a coalesced supernode labeled $0$, a bridge node between them labeled $1$, and the chain as a one-dimensional chain with length $L$ ($\ll N_c$) labeled $2$, $3$, $\dots$, $L+1$ that is attached to the bridge node.
We set $\mathsf{W}_{01}=p$ and $\mathsf{W}_{00}=1-p$ (i.e., a self-loop), and the degree of the bridge node as $k_b$ ($\geq 2$).
The transition probabilities are uniformly distributed to the adjacent nodes, making $\mathsf{W}_{10}=(k_b-1)/k_b$ and $\mathsf{W}_{12}=1/k_b$.
In this scheme, Eq.~\eqref{eq:SERW} can be re-written as
\begin{equation}\label{eq:SERW_lol}
\begin{aligned}
    \tilde{T}_x(r|l) &= \frac{1-r}{2} \tilde{T}_{x-1}(r|l) + \frac{1-r}{2} \tilde{T}_{x+1}(r|l) \\
    & + r \frac{N_c}{N_c+l} \tilde{T}_0(r|l) + r \frac{1}{N_c+l} \sum_{z=1}^{l} \tilde{T}_z(r|l) + 1,
\end{aligned}
\end{equation}
where $\tilde{T}_x(r|l) \equiv T_{x, L+1}\left(r|\mathcal{S}=\{0, 1, \dots, l\}\right)$, $x \in [2, l-1]$, and $l \in [2, L]$. 
The first and second terms indicate exploring the adjacent nodes, and the third and fourth terms indicate exploiting the supernode and the other previously visited nodes, respectively.
Solving Eq.~\eqref{eq:SERW_lol}, $\tilde{T}_{x}(r|l)$ is obtained by
\begin{equation}\label{eq:T_x(r|l)}
\begin{aligned}
    \tilde{T}_x(r|l) &= B(l) \alpha^{x-1} + D(l) \alpha^{-(x-1)} + G(r, l),
\end{aligned}
\end{equation}
where
\begin{equation}
\begin{aligned}
    G(r, l) &= \frac{N_c}{N_c+l}\tilde{T}_0(r|l) + \frac{1}{N_c+l}\sum_{z=1}^{l}\tilde{T}_z(r|l) + \frac{1}{r},
\end{aligned}
\end{equation}
with $\alpha = \left(1+\sqrt{1-(1-r)^2}\right)/(1-r)$.
The coefficients $B(l)$, $D(l)$, and $G(r, l)$ are determined by using the boundary conditions at $x=0$, $1$, and $l$, and $x=L$ and $l=L$; see the detailed derivation in Appendix~\ref{sec:AppendixC}.
Since the walker starts from the supernode $0$ with $\mathcal{S}=\{0, 1\}$, $\tilde{T}_0(r|1)$ is obtained by
\begin{equation}\label{eq:T_0(r|1)}
\begin{aligned}
    \tilde{T}_0(r|1) = \tilde{T}_0^{(1)}(r|1) + \sum_{z=1}^{L} \Delta_z(r),
\end{aligned}
\end{equation}
where $\tilde{T}_0^{(1)}(r|1)$ indicates $\tilde{T}_0(r|1)$ with $L=1$ and $\Delta_z(r)$ indicates how much time is additionally needed to visit node $z+1$ from node $z$ ($\geq 1$) with $r$.

As the simplest case, let us consider a clique network as the core with $p=1/N_c$ and $k_b=N_c+1$.
Then, $\tilde{T}_0^{(1)}(r|1)$ is given by
\begin{equation}\label{eq:T_0^(1)(r|1)_comp}
\begin{aligned}
    \tilde{T}_0^{(1)}(r|1) = \frac{1}{1-r}(N_c+1)^2.
\end{aligned}
\end{equation}
From Eq.~\eqref{eq:T_0^(1)(r|1)_comp}, we can see that $\tilde{T}_0^{(1)}(r|1)$ increases as $r$ increases. 
For large $N_c$, the ratio $\Delta_z(r)/\Delta_z(0)$ can be obtained by
\begin{equation}\label{eq:Delta_i(r)_comp}
\begin{aligned}
    \frac{\Delta_z(r)}{\Delta_z(0)} \simeq
    \begin{cases} 
    1 - \frac{1}{2} N_c z(z-1) r + \mathcal{O}\left(r^2\right) &\mbox{for } r \ll 1, \\
    \infty &\mbox{for } r \rightarrow 1,
    \end{cases}
\end{aligned}
\end{equation}
with 
\begin{equation}\label{eq:Delta_z(0)_comp}
\begin{aligned}
    \Delta_z(0)=N_c^2 + N_c - 1 + 2z.
\end{aligned}
\end{equation} 
Of course, $\Delta_z(r)/\Delta_z(0)$ goes to $1$ and $\infty$ in the limit $r \rightarrow 0$ and $r \rightarrow 1$, respectively.
We note that the first-order term of $\Delta_z(r)$ is proportional to $N_c^3$ whereas $\tilde{T}_0^{(1)}(r|1)$ is proportional to $N_c^2$.
Thus, the effect of exploitation is dominated by the $\Delta_z(r)$ term more than $\tilde{T}_0^{(1)}(r|1)$ as $N_c$ becomes larger.
When $0 < r \ll 1$ in Eq.~\eqref{eq:Delta_i(r)_comp}, $\Delta_z(r)/\Delta_z(0)$ decreases with increasing $r$, meaning that exploitation can help to explore the next node $z+1$ from node $z$.
Larger $N_c$ and $z$ induce smaller $\Delta_z(r)/\Delta_z(0)$, thus we can expect that $\tilde{T}_0(r|1)/\tilde{T}_0^{\rm NRW}$ also decreases with increasing $N_c$ and $L$ where $\tilde{T}_0^{\rm NRW} \equiv \tilde{T}_0(0|1)$, i.e., the MFPT of the NRW.

Figure~\ref{fig3}(b) shows $\tilde{T}_0(r|1)/\tilde{T}_0^{\rm NRW}$ at $L=2$ with different $N_c$.
As $N_c$ increases, $\tilde{T}_0(r|1)/\tilde{T}_0^{\rm NRW}$ has a steeper slope around $r=0$ and a smaller optimal value at $r^*$ which minimizes $\tilde{T}_0(r|1)$.
Here, the optimal ratio $\tilde{T}_0(r^*|1)/\tilde{T}_0^{\rm NRW}$ at $L=2$ converges to $1/2$ as $N_c \rightarrow \infty$ [inset in Fig.~\ref{fig3}(b)].
In Fig.~\ref{fig3}(c), we observe that $\tilde{T}_1(r|2)/\tilde{T}_0^{\rm NRW}$ also has a much deeper curve and the optimal ratio becomes smaller with increasing $L$.
From these results, we can conclude that exploitation gives advantages to finding distant nodes, and also that these advantages become more prominent as the size of the core and the distance to the target increase.
We note that the simulations in Fig.~\ref{fig3} are performed on a pre-coalesced network where $\mathcal{S}$ initially contains the starting node in the core, which suggests that our assumption in Eq.~\eqref{eq:SERW_lol} is valid.

Under what conditions would exploitation benefit a search in a lollipop-like network with a sparse core?
This question is equivalent to finding the condition $d\tilde{T}_0(r|1)/dr|_{r=0} < 0$, so we take account of the behavior of $\tilde{T}_0(r|1)$ when $r \ll 1$.
The occupation probability at $r \ll 1$ is almost the same as that of the NRW so that $p$ can be set as the occupation probability of the bridge node, i.e., $p \simeq (k_b-1)/N_c\langle k \rangle_c$ with the average degree of the core $\langle k \rangle_c$.
For large $N_c$, $\tilde{T}_0^{(1)}(r|1)$ is then given by
\begin{equation}\label{eq:T_0^(1)(r|1)_sparse}
\begin{aligned}
    \tilde{T}_0^{(1)}(r|1) \simeq \frac{N_c k_b \langle k \rangle_c}{k_b -1} \left( 1 - \frac{ \langle k \rangle_c - 2 k_b + 2 }{k_b - 1}r + \mathcal{O}\left(r^2\right) \right).
\end{aligned}
\end{equation}
Unlike the case in Eq.~\eqref{eq:T_0^(1)(r|1)_comp}, here $\tilde{T}_0^{(1)}(r|1)$ decreases when $\langle k \rangle_c > 2(k_b - 1)$ as $r$ increases. 
$\Delta_z(r)/\Delta_z(0)$ for large $z$ in the limit $N_c \rightarrow \infty$ is given by
\begin{equation}\label{eq:Delta_i(r)_sparse}
\begin{aligned}
    \frac{\Delta_z(r)}{\Delta_z(0)} \simeq
    1-\frac{(\langle k \rangle_c - 2) z^2}{2}r + \mathcal{O}\left(r^2\right),
\end{aligned}
\end{equation}
with 
\begin{equation}\label{eq:Delta_z(0)_sparse}
\begin{aligned}
    \Delta_z(0) = N_c \langle k \rangle_c + k_b + 2z-2.
\end{aligned}
\end{equation}
The slope of $\Delta_z(r)/\Delta_z(0)$ becomes steeper as $z$ increases, meaning that there exists a distance $L$ at which exploitation becomes beneficial when $\langle k \rangle_c > 2 $ is satisfied.
The minimum $E$ for a connected network is $N-1$, and in this case, $\langle k \rangle_c \simeq 2$ for large $N_c$, so that most networks except for the case of minimum $E$ can realize the benefit of exploitation when $L$ is large enough. 

%%%%%%%%%%%%% figure 4 %%%%%%%%%%%%%%
\begin{figure}[!t]
    \includegraphics[width=\columnwidth]{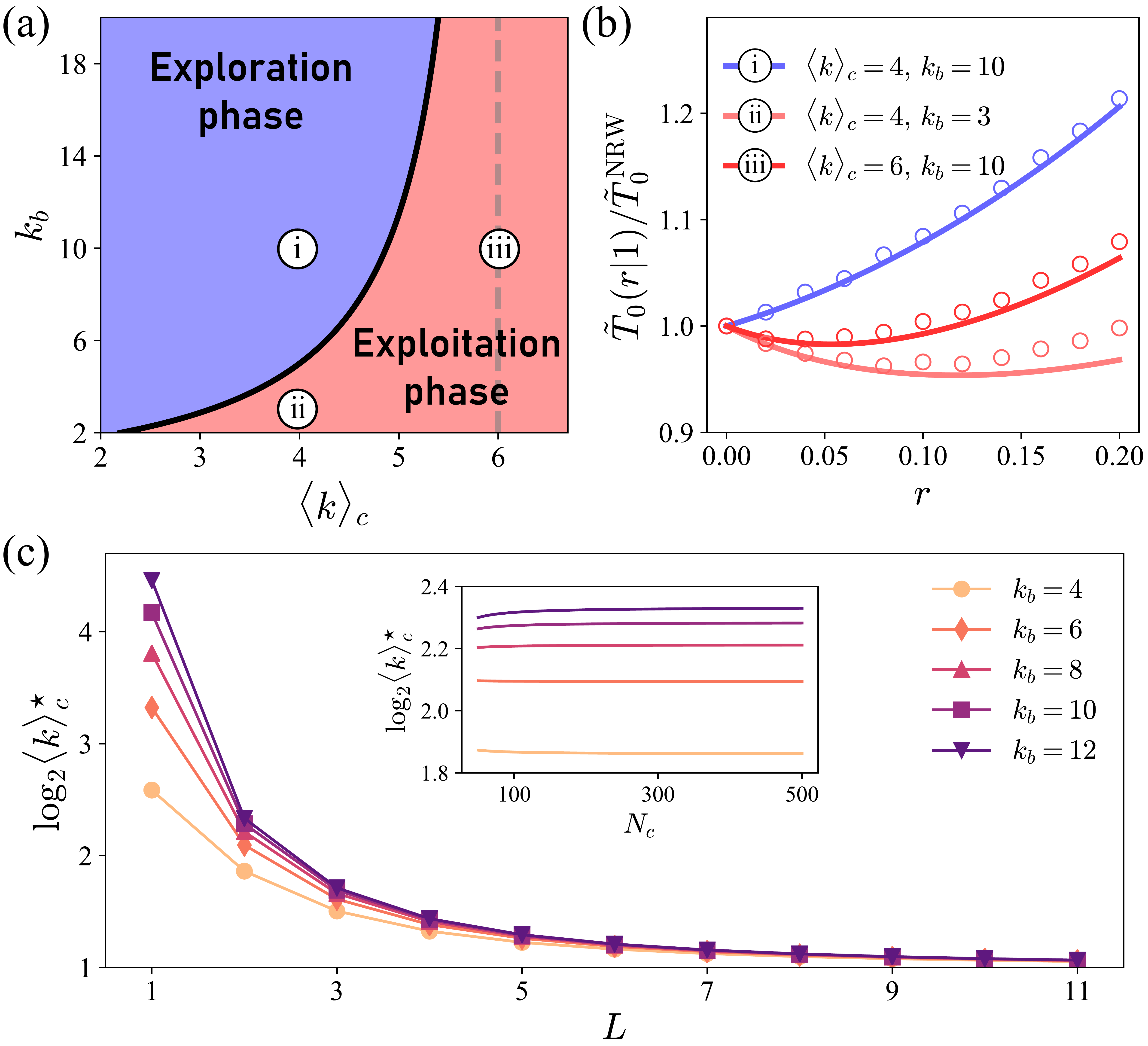}
    \vskip -0.1in
    \caption{ (a) Phase diagram of the SERW with $L=2$ at $0 < r \ll 1$. The axes correspond to the degree of the bridge node $k_b$ and the mean degree of the core $\langle k \rangle_c$.
    The exploitation (exploration) phase indicates the phase-space where exploitation can be beneficial (detrimental).
    The dashed line indicates the asymptotic value of $\langle k \rangle_c$.
    (b) Ratio between the MFPTs of the SERW and the NRW, denoted by $\tilde{T}_0(r|1)/\tilde{T}_0^{\rm NRW}$, as a function of the exploit probability $r$ with $p=(k_b-1)/N_c\langle k \rangle_c$, for the parameters indicated in (a). The symbols indicate the simulation results from $100$ ER networks with $N_c=400$. 
    (c) $\log_2\langle k \rangle_c^\star$ as a function of $L$ with different $k_b$, where $\langle k \rangle_c^\star$ indicates the phase boundary value of $\langle k \rangle_c$ at given $k_b$ and $L$. 
    The inset shows $\log_2\langle k \rangle_c^\star$ as a function of $N_c$.
    }\label{fig4}
\end{figure}
%%%%%%%%%%%%% figure 4 %%%%%%%%%%%%%%

Next, we identify a condition for beneficial exploitation at a particular $L$ and how the condition alters as $L$ increases.
Since obtaining a general form of the condition for $d \tilde{T}_0(r|1)/dr |_{r=0}< 0$ is difficult due to the complexity of Eq.~\eqref{eq:T_0(r|1)}, we firstly deal with the case of $L=2$ [the case of $L=1$ can be easily verified in Eq.~\eqref{eq:T_0^(1)(r|1)_sparse}].
Solving $d \tilde{T}_0(r|1)/dr |_{r=0}< 0$, the condition for large $N_c$ is given by
\begin{equation}\label{eq:D_2}
\begin{aligned}
    \langle k \rangle_c > \frac{(6k_b - 1) ( k_b - 1)}{k_b^2 - k_b - 1}.
\end{aligned}
\end{equation}
Figure~\ref{fig4}(a) shows a phase diagram where the exploitation (exploration) phase indicates the phase-space in which exploitation becomes beneficial (detrimental), as obtained from Eq.~\eqref{eq:D_2}.
The lower bound of $\langle k \rangle_c$ monotonically approaches $6$ as $k_b$ increases, and thus the range of $k_b$ for the exploitation phase widens as $\langle k \rangle_c$ increases. This also indicates that exploitation can provide advantages regardless of $k_b$ when $\langle k \rangle_c \geq 6$.
We plot three examples of $\tilde{T}_0(r|1)/\tilde{T}_0^{\rm NRW}$ as a function of $r$ in Fig.~\ref{fig4}(b), where (i) is in the exploration phase and (ii) and (iii) are in the exploitation phase.
There are slight differences from the simulation result as $r$ increases due to the assumption of $r\ll1$ in our analytic form, but the figure shows that our model well predicts which phase the network is in.

Figure~\ref{fig4}(c) illustrates how the phase boundary for $d \tilde{T}_0(r|1)/dr |_{r=0} = 0$ at large $N_c$ varies as $L$ increases. 
Here, $\langle k \rangle_c^\star$ denotes the phase boundary value of $\langle k \rangle_c$ at a given $k_b$ and $L$; when $\langle k \rangle_c > \langle k \rangle_c^\star$, a walker is in the exploitation phase.
We can observe that all $\langle k \rangle_c^\star$ with different $k_b$ monotonically decrease and converge to $2$ as $L$ increases, which is in agreement with the results from Eq.~\eqref{eq:Delta_i(r)_sparse}.
Another feature is decreasing the $\langle k \rangle_c^\star$ as $k_b$ decreases, implying that exploitation can be beneficial at smaller $\langle k \rangle_c$ as $k_b$ decreases.
The inset of Fig.~\ref{fig3}(c) shows that $\langle k \rangle_c^\star$ hardly changes as $N_c$ increases, implying that our analyses with $N_c \gg 1$ are mostly consistent for changing $N_c$. 

Lastly, we note that $C_0(r)$ is equal to $\tilde{T}_0(r|1)$ because a walker must visit nodes $0, 1, \dots, L$ to visit a node $L+1$.
But for the pre-coalesced lollipop-like network, the relation between the MCT and the MFPT should be clarified.
In this case, the MFPT from a node $i$ in the core to the end-node $T_{i, \rm end}(r)$ is much larger than the MFPTs between any other two nodes, indicating that the increasing rate of $\rho(t)$ is mainly dominated by $T_{i, \rm end}(r)$.
Using this fact, Eq.~\eqref{eq:C(r)} can be approximated as follows:
\begin{equation}\label{eq:CvsMFPT}
\begin{aligned}
    C_i(r) &\simeq \sum_{t=1}^{\infty} \left[1 - \left( 1-\exp{[-t/T_{i, {\rm end}}(r)]} \right) \right] \\
    &= \frac{1}{\exp{[1/T_{i, {\rm end}}(r)]}-1} \simeq T_{i, \rm end}(r),
\end{aligned}
\end{equation}
where $T_{i, {\rm end}}(r) \gg 1$.
In Appendix~\ref{sec:AppendixD}, we numerically confirm this linear relationship between the maximum MFPT and the GMFPT in the NRW in real-world networks.

\section{Real-world networks}
\label{sec:real_networks}

%%%%%%%%%%%%% figure 5 %%%%%%%%%%%%%%
\begin{figure}[!t]
    \includegraphics[width=\linewidth]{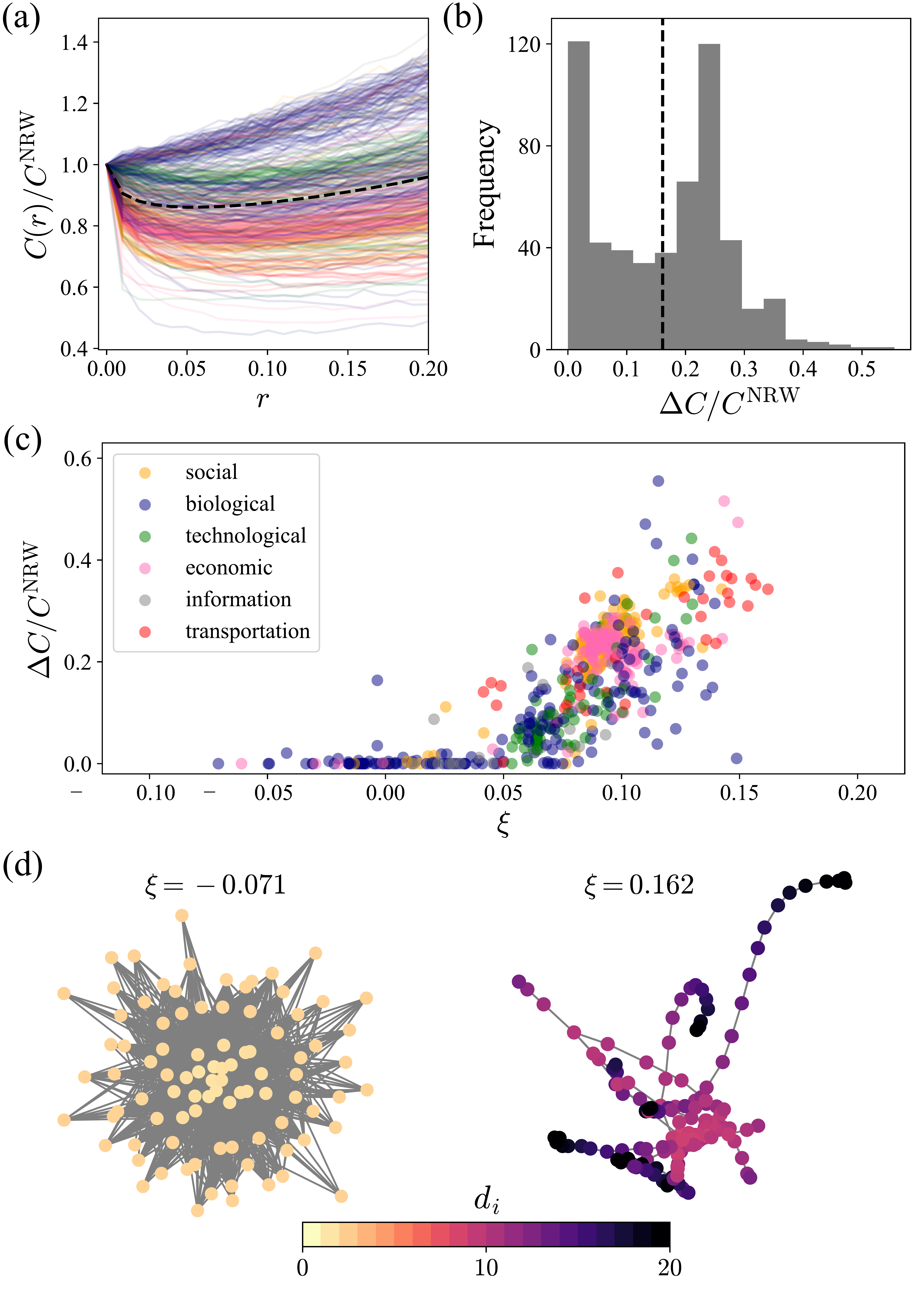}
    \vskip -0.1in
    \caption{
    (a) Ratio between the GMCTs of the SERW and the NRW, denoted $C(r)/C^{\rm NRW}$, as a function of $r$. The colored solid lines indicate the numerical results from $550$ real-world networks, and the dashed line indicates the average of the results. Line colors correspond to the network domains in (c).
    (b) Frequency histogram of $\Delta C/C^{\rm NRW}$ of the networks, where $\Delta C = C^{\rm NRW}-C(r^*)$ and the optimal exploit probability $r^*$ is numerically obtained by simulations. The dashed line indicates the average of $\Delta C/C^{\rm NRW}$ in the networks.
    (c) $\Delta C/C^{\rm NRW}$ vs. lollipop-likeness $\xi$.
    (d) Illustrations of real-world networks having the smallest (left) or the largest (right) $\xi$, respectively. Each node color displays the average distance from $i$ to others, denoted by $d_i$.
    }\label{fig5}
\end{figure}
%%%%%%%%%%%%% figure 5 %%%%%%%%%%%%%%

In this section, we apply the SERW to $550$ various real-world networks gathered from Ref.~\onlinecite{ghasemian2020stacking}.
These networks are divided into six main domains: social ($23\%$), biological ($32\%$), technological ($12\%$), economic ($23\%$), information ($3\%$), and transportation ($7\%$).
All networks are converted to undirected, unweighted, and connected networks before simulations.
Each simulation starts with a randomly selected starting node until the random walker visits every node in the network.
Estimating $C(r)$ from $1000$ simulations per network with different $r$ from $0$ to $0.2$, we verify whether exploitation can reduce the GMCT of the real-world networks.
As shown in Fig.~\ref{fig5}(a) and (b), numerous networks have an optimal exploit probability $r^*$, and we can even see that exploitation reduces the GMCT by about half in some networks.
Quantifying the extent that exploitation reduces GMCT by $\Delta C/C^{\rm NRW}$, where $\Delta C = C^{\rm NRW}-C(r^*)$, we can see that $75\%$ of all networks have advantages by $\Delta C/C^{\rm NRW} > 0.05$, and that the average of $\Delta C/C^{\rm NRW}$ is $0.161$.

Based on the results in the previous section, we conjecture that the more similar a network structure is to the lollipop structure, the more advantages it can gain by exploitation.
To quantify how similar a network is to a lollipop network, we focus on two main properties: (i) nodes in the chain have a larger average distance (i.e., shortest path length) than nodes in the core, and (ii) nodes in the chain (core) are connected to each other. 
Reflecting these properties, we define the lollipop-likeness $\xi$ as the multiplication of the Gini coefficient $g_{d}$ of average distances and the correlation coefficient $c_{d}$ between distances of each node and its neighbors' average distances (see details in Appendix~\ref{sec:AppendixD}).
For example, a clique network has nodes with the same distance, so $g_{d} = 0$ and $\xi = 0$. By contrast, $g_d$ of a lollipop network becomes larger as the chain lengthens, and $c_d$ is almost $1$ so that a lollipop network has a large $\xi$.

Figure~\ref{fig5}(c) shows a strong positive relation between $\xi$ and $\Delta C/C^{\rm NRW}$, with a Pearson (Spearman) correlation coefficient of $0.776$ ($0.786$) with the associated $p$ value less than $10^{-111}$ ($10^{-116}$).
At low $\xi$ [left panel in Fig.~\ref{fig5}(d)], nodes in the network generally have similar average distances which make $c_d < 0$.
At high $\xi$ [right panel in Fig.~\ref{fig5}(d)], some nodes form a chain-like structure, inducing large $g_d$ and $c_d$.
This positive relationship strongly supports our argument that exploitation becomes more beneficial as the network becomes more like a lollipop network.

We additionally check in which domains exploitation can be more helpful. 
As detailed in Table~\ref{tab2}, most networks in economic, social, and transportation domains satisfy $\Delta C/C^{\rm NRW} > 0.05$, whereas the information domain has a relatively lower fraction of networks satisfying the condition.
Interestingly, the orders of the fraction of networks having benefits by exploitation, $\langle \Delta C/C^{\rm NRW} \rangle$, and $\langle \Delta C/C^{\rm NRW} \rangle^*$ (the average of networks satisfying $\Delta C/C^{\rm NRW} > 0.05$) are almost same as the order of $\langle \xi \rangle$, which again supports the positive relation between $\xi$ and $\Delta C/C^{\rm NRW}$.
The distribution of $\xi$ for each domain is presented in Appendix~\ref{sec:AppendixD}.

\begin{table}[!t]
    \centering
    \caption{Averages of cover time benefit $\Delta C/C^{\rm NRW}$ and $\xi$ by domain in real-world networks. The number in parentheses denotes the number of networks included in the domain. We display the fraction of networks having benefits by exploitation, i.e., $\Delta C/C^{\rm NRW} > 0.05$ (Frac), the average of $\Delta C/C^{\rm NRW}$ in all networks ($\langle \Delta C/C^{\rm NRW}\rangle$), the average of $\Delta C/C^{\rm NRW}$ in networks having benefits by exploitation ($\langle \Delta C/C^{\rm NRW}\rangle^*$), and the ensemble average of lollipop-likeness $\xi$ ($\langle \xi \rangle$).}
    \label{tab2}
    \begin{tabular}{c|cccc}
    \noalign{\smallskip}\noalign{\smallskip}\hline\hline
    Domain &  Frac & $\langle \Delta C/C^{\rm NRW} \rangle$ & $\langle \Delta C/C^{\rm NRW} \rangle^*$ & $\langle \xi \rangle$ \\
    \hline
    Information (18) & 0.389 & 0.053 & 0.121 & 0.051 \\
    Biological (179) & 0.486 & 0.090 & 0.176 & 0.056 \\
    Technological (67)  & 0.761 & 0.120 & 0.151 & 0.080 \\
    Social (124) & 0.927 & 0.228 & 0.246 & 0.091 \\
    Economic (124) & 0.952 & 0.214 & 0.225 & 0.093 \\
    Transportation (38) & 0.895 & 0.241 & 0.248 & 0.107 \\
    \hline
    \hline
    \end{tabular}
\end{table}

\section{Conclusions}
\label{sec:conclusions}

Though a balance between exploitation and exploration is ubiquitous in various contexts, the effects of exploitation have not been fully clarified.
We revealed in this work that the impact of exploitation on target search varies according to the network structure.
To reflect long-term memory effects as many animals or organizations do, we built a non-Markovian model with a stochastic revisit to one of the previously visited nodes.
Two structures were analytically considered, a clique-like network and a lollipop-like network, and we showed that they have different behaviors with respect to the exploit probability.
Whereas exploitation hinders target search in the clique-like networks, it helps significantly reduce the MFPT and the MCT starting from the core in the lollipop-like networks.
We found that the benefit of exploitation in the latter becomes larger as the length of the chain increases, and from this, drew a phase diagram that depicts when exploitation can be beneficial.
Lastly, we verified that our model significantly reduces the GMCT in many real networks by simulations, and also revealed a strong association between the lollipop-likeness $\xi$ of a network and the benefit of exploitation $\Delta C/C^{\rm NRW}$.

We analytically and numerically proved that exploitation can give advantages to target search performance in complex networks; however, the effect of exploitation deserves further discussion. 
First, we assumed that the underlying network is static, but in many real situations like the spread of disease, the links between nodes can vary over time~\cite{holme2012temporal}.
Following reports that organisms dynamically alter the balance between exploitation and exploration as their environments change~\cite{daw2006cortical, cohen2007should}, it will be interesting to see how the effect of exploitation varies in temporal networks.
Second, we only dealt with a single random walker. In everyday observations though, various animals or organizations interact with each other, and this interdependence can affect the tension between exploration and exploitation.
Finally, we considered the case in which a walker exploits one of the previously visited nodes with equal probabilities. But in some contexts, walkers can have preferences so that the probabilities of exploitation may not be uniform~\cite{song2010modelling, boyer2014random}. It will also be intriguing to study how such heterogeneous exploitation probabilities could alter the effect of exploitation.

Despite these limitations, we expect that our work will give a hint as to why many organisms choose exploitation in decision-making contexts and also clarify how the advantages of exploitation depend on the underlying search spaces.
Furthermore, we proposed new directions for achieving efficient search performance in various non-Markovian models in the future.
In other words, search performance could be improved by introducing a strategy in which a walker goes to rather than simply avoids previously visited places.
A more efficient way to use the memory of past trajectories should be sufficiently discussed in future studies.

\begin{acknowledgments}
This study was supported by the Basic Science Research Program through the National Research Foundation of Korea (NRF Grant No. 2022R1A2B5B02001752).

\end{acknowledgments}

\appendix

\section{Derivation of Eq.~\eqref{eq:SERW}}
\label{sec:AppendixA}

Let $F_{ij}(t, r | \mathcal{S})$ define the first-passage time probability that a walker reaches $j$ starting from $i$ at $t$ steps with the exploit probability $r$ and a given set of distinct nodes $\mathcal{S}$ which includes $i$ at least.
For $i \neq j$, $F_{ij}(t, r | \mathcal{S})$ satisfies the backward equation:
\begin{equation}\label{App_eq:backward}
\begin{aligned}
    F_{ij}\left(t+1, r | \mathcal{S}\right) = \sum_{l=0}^{N-1} \mathsf{W}_{il}' F_{lj}\left(t, r | \mathcal{S}, X_0=i, X_{1}=l \right),
\end{aligned}
\end{equation}
where $X_0$ ($X_1$) indicates the current node (the next node) and the boundary condition $F_{jj}\left(t, r | \mathcal{S}\right)=\delta_{0, t}$.
Since the walker explores or exploits with exploit probability $r$, the transition probability is defined as $\mathsf{W}_{il}' \equiv (1-r)\mathsf{W}_{il} + r \mathbbm{1}_{\mathcal{S}}(l)/|\mathcal{S}|$ where $\mathbbm{1}_{\mathcal{S}}(l)$ denotes the indicator function, i.e., $\mathbbm{1}_{\mathcal{S}}(l)$ equals to $1$ if $l \in \mathcal{S}$ and $0$ otherwise. Plugging the definition of $\mathsf{W}_{il}'$ into Eq.~\eqref{App_eq:backward}, the backward equation can be expressed by
\begin{equation}\label{App_eq:backward2}
\begin{aligned}
    F_{ij}\left(t+1, r | \mathcal{S}\right) = (1-r) \sum_{l=0}^{N-1} \mathsf{W}_{il} F_{lj}\left(t, r | \mathcal{S}' \right)  
    \\+ \frac{r}{|\mathcal{S}|} \sum_{l\in \mathcal{S}} F_{lj}\left(t, r | \mathcal{S}\right),
\end{aligned}
\end{equation}
where $\mathcal{S}' \equiv \mathcal{S} \cup l$.
The first term in Eq.~\eqref{App_eq:backward2} describes the exploration process, and we used the relation $F_{lj}\left(t, r | \mathcal{S}, X_0=i, X_{1}=l \right) = F_{lj}(t, r|\mathcal{S}')$. 
In the second term, which represents the exploitation process, the node $l$ is not a newly visited node but randomly chosen in $\mathcal{S}$, leading to the relation $F_{lj}(t, r|\mathcal{S}') = F_{lj}(t, r|\mathcal{S})$. Using Eq.~\eqref{App_eq:backward2} and the definition of MFPT from $i$ to $j$, $T_{ij}(r|\mathcal{S}) \equiv \sum_{t=0}^{\infty}t F_{ij}(t, r|\mathcal{S})$, we obtain
\begin{equation}\label{app_eq:SERW}
\begin{aligned}
    T_{ij}\left(r | \mathcal{S}\right) = (1-r) \sum_{l=0}^{N-1}\mathsf{W}_{il} T_{lj} \left(r | \mathcal{S}'\right) \\
    + \frac{r}{|\mathcal{S}|}\sum_{l \in \mathcal{S}} T_{lj}\left( r|\mathcal{S} \right) + 1,
\end{aligned}
\end{equation}
with the boundary condition $T_{jj}(r|\mathcal{S})=0$.

\section{Clique networks}
\label{sec:AppendixB}

Regardless of which node a walker starts from, in the case of a clique network of $N$ nodes, the probability $p_z$ of finding the $(z+1)$th new node is
\begin{equation}\label{app_eq:pi}
\begin{aligned}
    p_z = (1-r) \frac{N-z}{N-1},
\end{aligned}
\end{equation}
with $i=z, \dots, N-1$.
When $r=0$, Eq.~\eqref{app_eq:pi} becomes almost equivalent to the probability of collecting a new coupon in the coupon collector problem~\cite{maziya2020dynamically}. The MCT $C_i$ is then represented by the sum of individual waiting times $1/p_z$ between the $z$th and $(z+1)$th new node:
\begin{equation}\label{app_eq:C_comp}
\begin{aligned}
    C_i (r) &= \frac{N-1}{1-r}\sum_{z=1}^{N-1} \frac{1}{N-z} \\
    &\simeq \frac{N-1}{1-r} \left[ \ln{(N-1)} + \gamma \right], \;\; {\rm for} \; N \gg 1,
\end{aligned}
\end{equation}
where $\gamma = 0.5772\dots$ is the Euler--Mascheroni constant. 
For the NRW, the clique network provides the lower bound of the cover time given by $\left(1+o(1)\right) N \ln N$.

\begin{figure}[!t]
    \includegraphics[width=\columnwidth]{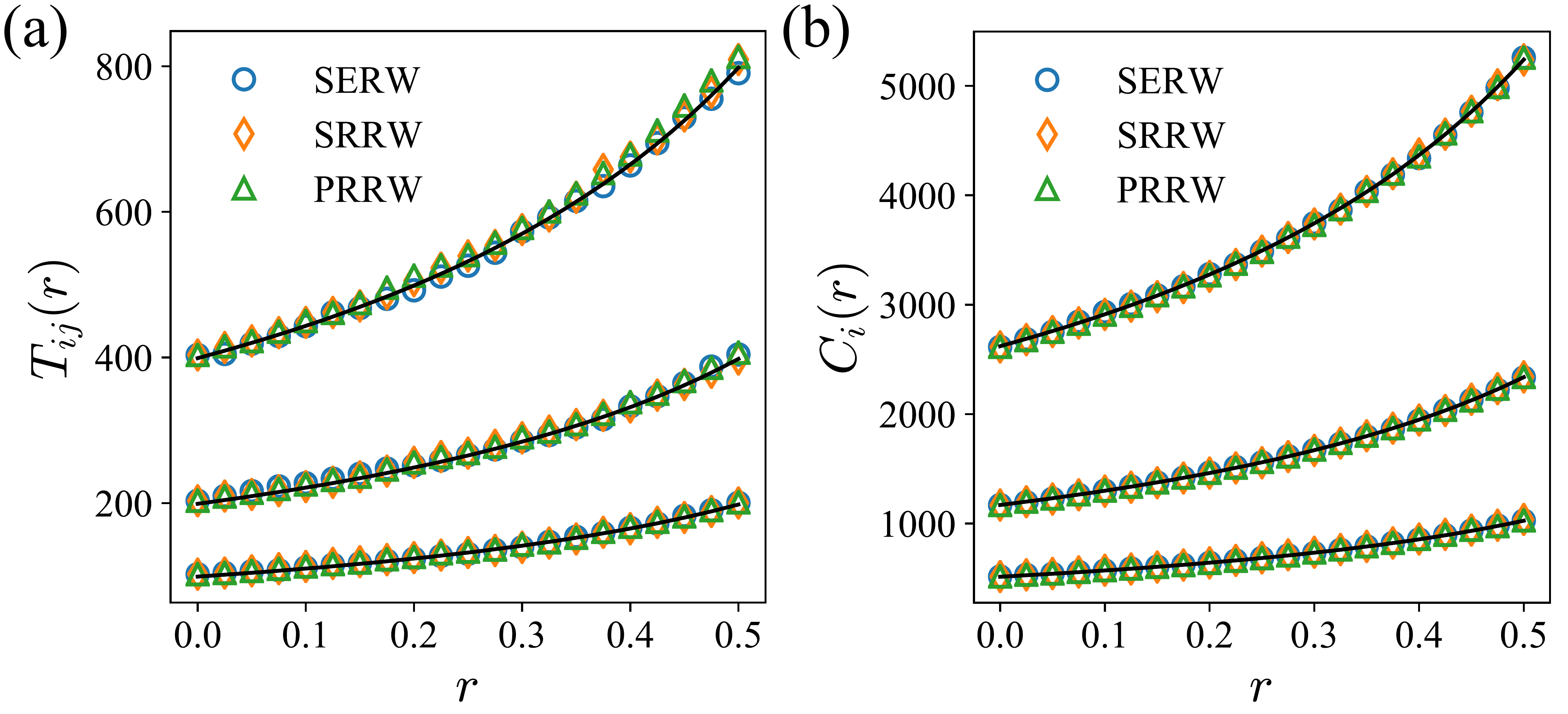}
    \vskip -0.1in
    \caption{ (a) The MFPT $T_{ij}(r)$ from $i$ to $j$ and (b) the MCT $C_i(r)$ in a clique network as a function of exploit probability $r$ with different random walks with returns: stochastic exploiting random walk (SERW), stochastic resetting random walk (SRRW), and preferential relocating random walk (PRRW).
    From bottom to top in both panels, the results are obtained for a clique network with $N=100$, $200$, and $400$.
    Solid lines (symbols) indicate the analytical (simulation) results.
    }\label{AppFig1}
\end{figure}

To derive Eq.~\eqref{app_eq:C_comp} from Eq.~\eqref{eq:FPT_exp(t)}, let us calculate the GMFPT for a target node $v$ from Eq.~\eqref{eq:SERW}. In a clique network, all nodes are symmetric so that $T_{lj}\left( r|\mathcal{S} \right)$ are identical for any node $l$ in $\mathcal{S}$. Applying $\mathsf{W}_{il} = 1/(N-1)$, Eq.~\eqref{app_eq:C_comp} can be rewritten as 
\begin{equation}\label{app_eq:SERW_comp}
\begin{aligned}
    T_{ij}\left(r  | \mathcal{S} \right) = (1-r) \frac{1}{N-1} \sum_{l \neq i} T_{lj} \left(r  | \mathcal{S}' \right)
    + r T_{ij}\left( r | \mathcal{S} \right) + 1.
\end{aligned}
\end{equation}
Since $T_{ij}\left(r  | \mathcal{S} \right)$ is independent of $\mathcal{S}$ if $i \in \mathcal{S}$, we can reduce $T_{ij}(r|\mathcal{S})$ to $T_{ij}(r)$, which can be obtained by 
\begin{equation}
\begin{aligned}
    T_{ij}(r) = \frac{N-1}{1-r},
\end{aligned}
\end{equation}
for any pair of nodes $i$ and $j$. Since the MFPT of any pairs in the clique network are the same as $T_j^{\rm NRW} = N-1$, the GMFPT $T_j(r)$ follows $T_j(r) = T_j^{\rm NRW}/(1-r)$. 
Plugging $T_j(r)$ into the integral form in Eq.~\eqref{eq:C(r)}, $C(r)$ for a clique network is calculated by
\begin{equation}
\begin{aligned}
    C(r) &= \int_{0}^{\infty} dt \left[1-\{1-\exp{\left(-t/T_j(r)\right)}\}^{N-1}\right]
    \\&= T_j(r)\sum_{i=1}^{N-1} \frac{1}{N-i},
\end{aligned}
\end{equation}
which is the same result as Eq.~\eqref{app_eq:C_comp}. 
As we mentioned in the derivations, both $T_{ij}(r)$ and $C(r)$ have no dependence on $\mathcal{S}$, hence our results for clique networks are also valid in other random walks with returns, e.g., stochastic resetting random walks or preferential relocating random walks. Figure~\ref{AppFig1} shows that the MFPT and the MCT of these three models with $r$ are identical in a clique network.

%%%%%%%%%%%%% Appendix figure 1 %%%%%%%%%%%%%%
%%%%%%%%%%%%% Appendix figure 1 %%%%%%%%%%%%%%

\section{Lollipop-like networks}
\label{sec:AppendixC}

To derive Eq.~\eqref{eq:T_0(r|1)}, we rewrite Eq.~\eqref{eq:SERW_lol}:
\begin{equation}\label{app_eq:SERW_lol}
\begin{aligned}
    \tilde{T}_x(r|l) &= \frac{1-r}{2} \tilde{T}_{x-1}(r|l) + \frac{1-r}{2} \tilde{T}_{x+1}(r|l) \\
    & + r \frac{N_c}{N_c+l} \tilde{T}_1(r|l) + r \frac{1}{N_c+l} \sum_{i=2}^{l} \tilde{T}_i(r|l) + 1,
\end{aligned}
\end{equation}
for $2 \leq x \leq l-1$ and $2 \leq l \leq L$.
Solving Eq.~\eqref{app_eq:SERW_lol} yields
\begin{equation}\label{app_eq:T_x(r|l)}
\begin{aligned}
    \tilde{T}_x(r|l) &= B(l) \alpha^{x-1} + D(l) \alpha^{-(x-1)} + G(r, l),
\end{aligned}
\end{equation}
with the definition
\begin{equation}\label{app_eq:G(l)}
\begin{aligned}
    G(r, l) &=  \frac{N_c}{N_c+l}\tilde{T}_0(r|l) + \frac{1}{N_c+l}\sum_{i=1}^{l}\tilde{T}_i(r|l) + \frac{1}{r},
\end{aligned}
\end{equation}
and $\alpha = \left(1+\sqrt{1-(1-r)^2}\right)/(1-r)$.
There exist four boundary conditions at (i) $x=0$, (ii) $x=1$, (iii) $x=l$, and (iv) $x=L+1$ and $l=L+1$, which are given by
\begin{equation}\label{app_eq:SERW_lol_BC1}
\begin{aligned}
    {\rm (i)}\, \tilde{T}_0(r|l) &= (1-r)(1-p) \tilde{T}_{0}(r|l) + (1-r)p \tilde{T}_{1}(r|l) \\
    & + r G(r, l),\\
    {\rm (ii)}\, \tilde{T}_1(r|l) &= (1-r)\frac{k_b-1}{k_b} \tilde{T}_{0}(r|l) + (1-r)\frac{1}{k_b} \tilde{T}_{2}(r|l) \\
    & + r G(r, l),\\
    {\rm (iii)}\, \tilde{T}_l(r|l) &= \frac{1-r}{2} \tilde{T}_{l-1}(r|l) + \frac{1-r}{2} \tilde{T}_{l+1}(r|l+1) \\
    & + r G(r, l),
\end{aligned}
\end{equation}
for $l \geq 2$ or
\begin{equation}\label{app_eq:SERW_lol_BC2}
\begin{aligned}
    {\rm (iii)}\, \tilde{T}_1(r|1) &= (1-r)\frac{k_b-1}{k_b} \tilde{T}_{0}(r|1) + (1-r)\frac{1}{k_b} \tilde{T}_{2}(r|2) \\
    & + r G(r, 1),\\
\end{aligned}
\end{equation}
for $l = 1$, and (iv) $T_{L+1}(r|L+1) = 0$.
Inserting Eq.~\eqref{app_eq:T_x(r|l)} into condition (i), we get
\begin{equation}\label{app_eq:(i)}
\begin{aligned}
    \tilde{T}_0(r|l) = \frac{1}{J} \left( 2p \tilde{T}_1(r|l) + \frac{\alpha}{(\alpha-1)^2} G(r, l) \right).
\end{aligned}
\end{equation}
Here, $J = (\alpha-1)^2/\alpha + 2p $. Since we need to obtain $\tilde{T}_0(r|l)$, the question is converted into how to find $\tilde{T}_1(r|l)$ and $G(r, l)$.
Substituting Eq.~\eqref{app_eq:(i)} for $\tilde{T}_0(r|l)$ in Eq.~\eqref{app_eq:G(l)} and condition (ii), we get
\begin{equation}\label{app_eq:BlDl1}
\begin{aligned}
    \left( \frac{2p}{J} + \frac{\alpha^{l}-1}{\alpha-1} \right) & B(l) + \left( \frac{2p}{J} + \frac{\alpha - \alpha^{-(l-1)}}{\alpha-1} \right) D(l) \\ &= -\frac{\alpha^2 + 1}{(\alpha-1)^2}(N_c +l),
\end{aligned}
\end{equation}
and
\begin{equation}\label{app_eq:BlDl2}
\begin{aligned}
    &\left( \alpha + \frac{1}{\alpha} - \frac{4p(k_b-1)}{k_bJ} - \frac{2\alpha}{k_b} \right) B(l) \\
    &+\left( \alpha + \frac{1}{\alpha} - \frac{4p(k_b-1)}{k_bJ} - \frac{2}{k_b \alpha} \right) D(l) = 0.
\end{aligned}
\end{equation}
Then by combining Eq.~\eqref{app_eq:BlDl1} and Eq.~\eqref{app_eq:BlDl2}, we can determine the expressions of $B(l)$ and $D(l)$.

To deduce the expression of $G(r, l)$, we apply condition (iii) in Eq.~\eqref{app_eq:SERW_lol_BC1} and get the relation
\begin{equation}\label{app_eq:G(l+1)-G(l)}
\begin{aligned}
    G(r, l+1) - G(r, l) &= - \left( B(l+1) - B(l) \right) \alpha^{l} \\
    &- \left( D(l+1) - D(l) \right) \alpha^{-l},
\end{aligned}
\end{equation}
for $l \geq 2$. 
Summing Eq.~\eqref{app_eq:G(l+1)-G(l)} from $l=1$ to $l=L$ by applying Eq.~\eqref{app_eq:SERW_lol_BC2} and condition (iv), we can obtain the expression of $G(r, 1)$ as
\begin{equation}\label{app_eq:G(1)}
\begin{aligned}
    G(r, 1) &= - \frac{k}{2} \left( \alpha + \frac{1}{\alpha} - \frac{4p(k_b-1)}{k_bJ} \right) (B(1) + D(1)) \\
    &+ \sum_{z=2}^{L} (\alpha -1 ) \left( D(z) \alpha^{-z} - B(z) \alpha^{z-1} \right).
\end{aligned}
\end{equation}

From the above results, we can obtain the analytic expression of $\tilde{T}_0(r|1)$ with $L$. 
For $L=1$, the MFPT from node $0$ to node $2$ becomes 
\begin{equation}\label{app_eq:T_0^(1)(r|1)}
\begin{aligned}
    \tilde{T}^{(1)}_0(r|1) &= -\frac{k_b}{2}\left( \alpha + \frac{1}{\alpha} - \frac{4p}{J} \right) \left( B(1) + D(1) \right),
\end{aligned}
\end{equation}
where $\tilde{T}^{(1)}_0(r|1)$ denotes $\tilde{T}_0(r|1)$ with $L=1$.
Next, for $L \geq 2$, plugging Eq.~\eqref{app_eq:T_x(r|l)} and Eq.~\eqref{app_eq:G(1)} into Eq.~\eqref{app_eq:(i)} with $l=1$, $\tilde{T}_0(r|1)$ can be represented by
\begin{equation}\label{app_eq:T_0(r|1)}
\begin{aligned}
    \tilde{T}_0(r|1) = \tilde{T}_0^{(1)}(r|1) + \sum_{z=1}^{L} \Delta_z(r),
\end{aligned}
\end{equation}
with
\begin{equation}\label{app_eq:Delta_z(r)}
\begin{aligned}
    \Delta_z(r) = (\alpha -1 ) \left( D(z) \alpha^{-z} - B(z) \alpha^{z-1} \right).
\end{aligned}
\end{equation}
Subtracting $\tilde{T}_0(r|1)$ with $L=L'$ from $\tilde{T}_0(r|1)$ with $L = L'+1$ leaves $\Delta_L'(r)$, so we can see that $\Delta_z(r)$ means the time a walker needs to visit node $z+1$ from node $z$.

Let us consider a lollipop network consisting of a clique with $qN$ nodes and a chain with $(1-q)N$ nodes with $q \in [0, 1]$. 
When $r=0$, we have $\tilde{T}_0^{(1)}(0|1) = qN (1+qN)$ and $\Delta_z(0)=2z+q^2N^2 -2$. Inserting them into Eq.~\eqref{app_eq:T_0(r|1)}, we obtain
\begin{equation}\label{app_eq:T_0(0|1)_lollipop}
\begin{aligned}
    \tilde{T}_0(0|1) = N \left( (1-q)q^2 N^2 + (1-q)^2 N + 2q - 1 \right).
\end{aligned}
\end{equation}
For large $N$, $\tilde{T}_0(0|1) \simeq (1-q)q^2 N^3$ and it is maximized to $\tilde{T}_0(0|1) \simeq 4N^3/27$ at $q=2/3$. 
This result is in agreement with Ref.~\onlinecite{brightwell1990maximum}, where they additionally showed that the lollipop network is extremal with respect to the MFPT.

\section{Real-world networks}
\label{sec:AppendixD}

%%%%%%%%%%%%% Appendix figure 2 %%%%%%%%%%%%%%
\begin{figure}[!t]
    \includegraphics[width=\columnwidth]{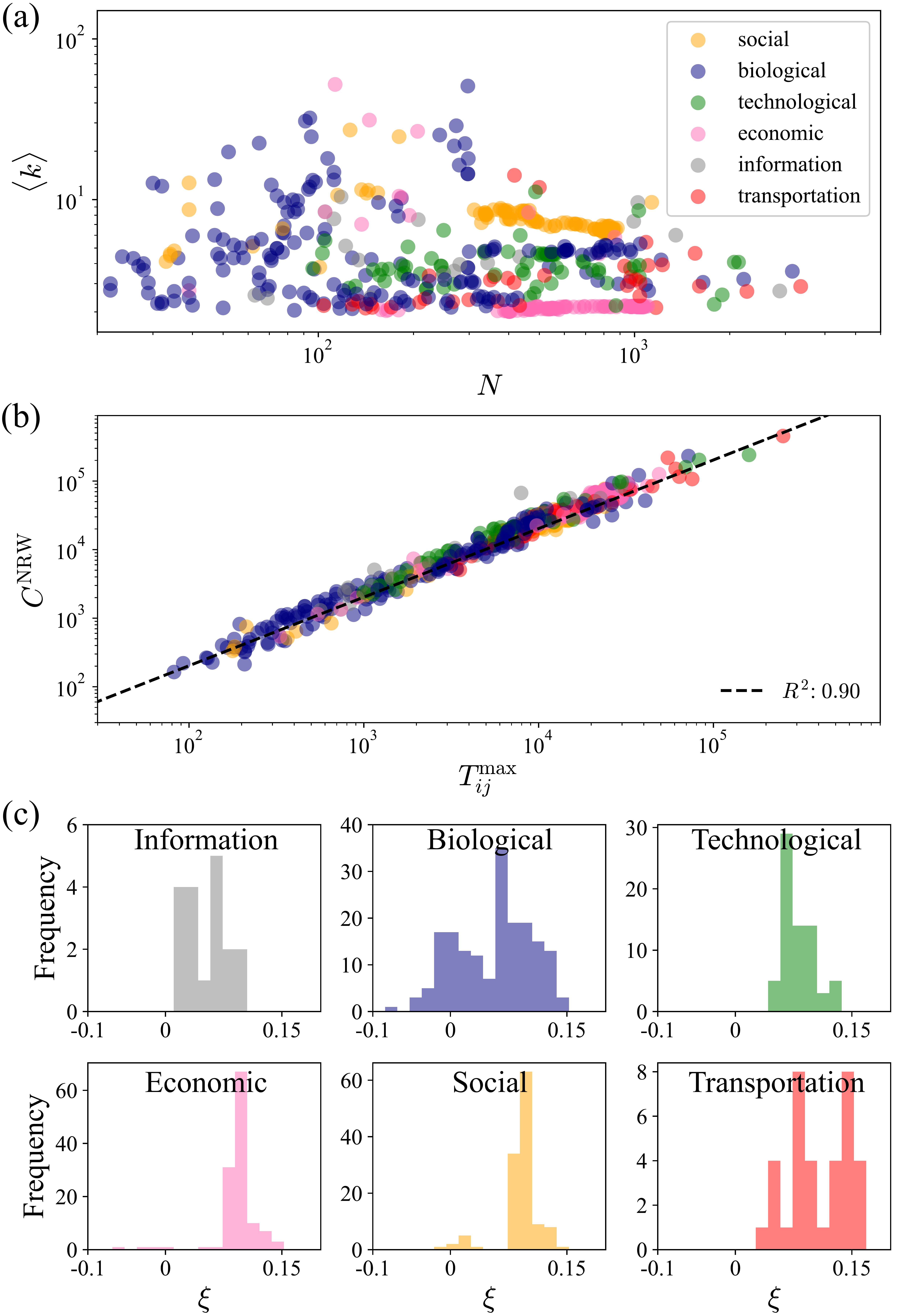}
    \vskip -0.1in
    \caption{ 
    (a) The number of nodes $N$ vs. the average of degree $k$ of real-world networks. 
    (b) The MCT of the NRW $C^{\rm NRW}$ vs. the maximum MFPT $T_{ij}^{\rm max}$. 
    The dashed line is the fitted line between $C^{\rm NRW}$ and $T_{ij}^{\rm max}$ with $R^2=0.90$ and $p \leq 10^{-274}$ by linear regression.
    (c) The distribution of lollipop-likeness $\xi$ in the six domains. The colors in (b) and (c) correspond to (a).
    }\label{AppFig2}
\end{figure}
%%%%%%%%%%%%% Appendix figure 2 %%%%%%%%%%%%%%

We use the $550$ diverse real-world network dataset from Ref.~\onlinecite{ghasemian2020stacking}. 
The networks are divided into six main domains with different numbers of networks, as follows: social ($124$), biological ($179$), economic ($124$), technological ($67$), information ($18$), and transportation ($38$); see Fig.~\ref{AppFig2}(a) or the supplementary information of Ref.~\onlinecite{ghasemian2020stacking} for plots of the average degree vs. number of nodes.

As mentioned in Sec.~\ref{sec:serw} B, we plot the maximum MFPT $T_{ij}^{\rm max}$ vs. $C^{\rm NRW}$ in Fig.~\ref{AppFig2}(b). 
As expected from Eq.~\eqref{eq:CvsMFPT}, we observe a strongly linear relation between $T_{ij}^{\rm max}$ and $C^{\rm NRW}$ with $R^2=0.90$.
The slope of the fitted line is not $1$ but $2.03$, which is thought to be because there is not one but several nodes with maximum MFPT.
We hope that follow-up studies will be performed to more clearly grasp the relationship between the FPT and CT quantities.

The lollipop-likeness $\xi$ is defined as $\xi \equiv g_d c_d$, where $g_d$ indicates the Gini coefficient of the average distances and $c_d$ indicates the correlation coefficient between the average distance of nodes and the mean of their neighbor's average distance.
$g_d$ and $c_d$ are calculated by
\begin{equation}
\begin{aligned}
    g_d = \frac{\sum_{i=0}^{N-1} \sum_{j=0}^{N-1} \left|d_i - d_j\right|}{2N^2 \overline{d}},
\end{aligned}
\end{equation}
and 
\begin{equation}
\begin{aligned}
    c_d = \frac{\sum_{i=0}^{N-1} \left(d_i - \overline{d}\right)\left( d_{i, \rm{nb}} - \overline{d}_{{\rm nb}}\right)}{\sqrt{\sum_{i=0}^{N-1} (d_i - \overline{d})^2} \sqrt{\sum_{j=0}^{N-1} (d_{i, {\rm nb}} - \overline{d}_{\rm nb} )^2}},
\end{aligned}
\end{equation}
where $d_i$ is the average distance from $i$ to others and $\overline{d} \equiv \sum_{i=0}^{N-1}d_i/N$.
The mean of average distance of neighbors is defined by $d_{i, {\rm nb}} \equiv \sum_{j=0}^{N-1} d_{j} \mathsf{A}_{ij}/k_i$ and $\overline{d}_{\rm nb} \equiv \sum_{i=0}^{N-1} d_{i, {\rm nb}} / N$.
Figure~\ref{AppFig2}(c) shows the distribution of $\xi$ in each domain. 
Here, if we look at the domains with large $\langle \xi \rangle$, the distributions of the economic and social domains are similar, while the transportation domain with the largest $\langle \xi \rangle$ has two peaks in its distribution. Here, the lower and upper peaks mainly consist of roads and public transport networks, respectively, because the nodes of public transport networks represent bus stops or subway stations which are often connected in series.

\section{Numerical calculations}
\label{sec:AppendixE}

As a direct summing up to $t=\infty$ is numerically impossible, we set the upper bound $t_{\rm cut}$ that satisfies $1-\rho(t_{\rm cut}) \leq 10^{-10}$, as in Ref.~\onlinecite{maier2017cover}. Then, Eq.~\eqref{eq:C(r)} can be written as
\begin{equation}
\begin{aligned}
    C(r) &= \sum_{t=1}^{\infty} [1-\rho(t)] \\
    &= \sum_{t=1}^{t_{\rm cut}} \left[ 1-\rho(t) \right] + \sum_{t=t_{\rm cut}+1}^{\infty} \left[ 1-\rho(t) \right].
\end{aligned}
\end{equation}
The first term is used for calculating $C(r)$ in Fig.~\ref{fig2}, and the second term indicates the error that emerges from the cutoff. 
Since $t_{\rm cut}$ is larger than $T_j$ for all $j$ to satisfy $1-\rho(t_{\rm cut}) \leq 10^{-10}$, $T^{\rm max}$ dominates $\rho(t)$.
Thus, the second term, denoted by $\mathcal{E}$, can be approximated by
\begin{equation}
\begin{aligned}
    \mathcal{E} &\simeq \int_{t_{\rm cut}}^{\infty} dt \left[1 - \left( 1-\exp{[-t/T^{\rm max}]} \right)^{n_{\rm max}} \right] \\
    &\simeq \int_{t_{\rm cut}}^{\infty} dt \left[1 - \left( 1- n_{\rm max} \exp{[-t/T^{\rm max}]} \right) \right] \\
    &= n_{\rm max} T^{\rm max} \exp{\left[-t_{\rm cut}/T^{\rm max} \right]}  \\
    &\leq n_{\rm max} T^{\rm max} \times 10^{-10},
\end{aligned}
\end{equation}
where $n_{\rm max}$ is the number of nodes whose $T_j$ is the same as $T^{\rm max}$.
In all the results we obtained in this paper, the conditions $T^{\rm max} \ll 10^{10}$ and $n_{\rm max} \sim \mathcal{O}(1)$ yield $\mathcal{E} \ll 1$.

\bibliography{main}

\end{document}